\documentclass[10pt, conference]{IEEEtran}
\usepackage{cite}
\usepackage{amsmath,amssymb,amsfonts}
\usepackage{algorithmic}
\usepackage{graphicx}
\usepackage{textcomp}
\usepackage[dvipsnames]{xcolor}
    
\usepackage[hidelinks]{hyperref}
\usepackage{mathtools}
\usepackage{bbm}
\usepackage{algorithmic, algorithm}
\usepackage{etoolbox}
\usepackage[top=0.7in,bottom=0.95in, inner = 0.7in, outer = 0.7in]{geometry}
\AtBeginEnvironment{bmatrix}{\setlength{\arraycolsep}{2pt}}

\usepackage{glossaries}
\glsdisablehyper
\newacronym{3gpp}{3GPP}{3rd Generation Partnership Project}
\newacronym{bs}{BS}{base station}
\newacronym{ue}{UE}{user equipment}
\newacronym{mdt}{MDT}{minimization of drive tests}
\newacronym{urllc}{URLLC}{ultra-reliable low-latency communications}
\newacronym{rss}{RSS}{received signal strength}
\newacronym[longplural={Gaussian processes }]{gp}{GP}{Gaussian processes}
\newacronym{csi}{CSI}{channel state information}
\newacronym{iot}{IoT}{internet of things}
\newacronym{pdf}{PDF}{probability density function}
\newacronym{cdf}{CDF}{cumulative distribution function}
\newacronym{snr}{SNR}{signal-to-noise ratio}
\newacronym{sinr}{SINR}{signal-to-interference-plus-noise ratio}
\newacronym{los}{LOS}{line of sight}
\newacronym{nlos}{NLOS}{non-line of sight}
\newacronym{tdma}{TDMA}{time-division multiple access}
\newacronym{3d}{3D}{three-dimensional}
\newacronym{pcr}{PCR}{probably correct reliability}

\newcommand{\mbf}[1]{\mathbf{#1}}
\newcommand{\mbs}[1]{\boldsymbol{#1}}

\newcommand*{\cond}{\hspace*{1pt} |\hspace*{1pt}}
\newcommand{\norm}[1]{\left\lVert#1\right\rVert}

\newcommand{\T}{\textsf{T}}

\DeclareMathOperator{\Cov}{Cov}
\DeclareMathOperator*{\argmax}{arg\,max}

\DeclareMathOperator*{\maximize}{maximize}

\begin{document}

\title{Predictive Rate Selection for Ultra-Reliable Communication using Statistical Radio Maps}

\author{\IEEEauthorblockN{Tobias~Kallehauge\IEEEauthorrefmark{1}, Pablo~Ramírez-Espinosa\IEEEauthorrefmark{1}, Anders~E.~Kalør\IEEEauthorrefmark{1}, Christophe~Biscio\IEEEauthorrefmark{2} and Petar~Popovski\IEEEauthorrefmark{1}}
\IEEEauthorblockA{\IEEEauthorrefmark{1}Department of Electronic Systems, Aalborg University, Denmark}
\IEEEauthorblockA{\IEEEauthorrefmark{2}Department of Mathematical Sciences, Aalborg University, Denmark\\
Email: \IEEEauthorrefmark{1}\{tkal,pres,aek,petarp\}@es.aau.dk \IEEEauthorrefmark{2}christophe@math.aau.dk}
}

\maketitle

\begin{abstract}
This paper proposes exploiting the spatial correlation of wireless channel statistics beyond the conventional received signal strength maps by constructing \textit{statistical radio maps} to predict any relevant channel statistics to assist communications. Specifically, from stored channel samples acquired by previous users in the network, we use Gaussian processes (GPs) to estimate quantiles of the channel distribution at a new position using a non-parametric model. This prior information is then used to select the transmission rate for some target level of reliability. The approach is tested with synthetic data, simulated from urban micro-cell environments, highlighting how the proposed solution helps to reduce the training estimation phase, which is especially attractive for the tight latency constraints inherent to ultra-reliable low-latency (URLLC) deployments. 
\end{abstract}


\section{Introduction} \label{sec:intro}

\Gls{urllc} is one of the most significant novelties brought by 5G, aiming to support wireless connections with very stringent requirements in terms of latency and reliability. From a physical layer viewpoint, the inherent randomness of wireless channels is the major challenge, rendering a fundamentally unreliable communication link. In cellular systems, instantaneous \gls{csi} estimation is widely used to adapt the resource allocation and transmission rate to the actual channel state, requiring an estimation phase before the data transmission. However, when looking towards \gls{urllc}, two main drawbacks arise: \textit{i)} it introduces a latency that may be unacceptable for mission-critical applications; and \textit{ii)} in the ultra-reliable regime, assuming the channel stays \textit{exactly} constant during both estimation and transmission phases may be wrong \cite{Swamy2019}, depending on the actual channel dynamics.

As an alternative, the communication parameters can be chosen according to  \textit{statistical} knowledge of the channel, based on parametric or non-parametric models, so that a reliability constraint is met with given confidence \cite{Angjelichinoski2019}. While parametric models can be fitted using relatively few samples, the models widely used in wireless communications, particularly  Rayleigh or Rician, are not intended for \gls{urllc}~\cite{Eggers2019}. Models based on, e.g., extreme value theory, can be better suited for \gls{urllc}, but they can also be challenging to apply in practice as they rely on results that are valid only asymptotically~\cite{Mehrnia2022}.
At the other extreme are non-parametric models, which palliate the issue of model mismatch but require a massive number of samples that is prohibitive in most realistic situations, especially for non-stationary environments \cite{Angjelichinoski2019}.

Reducing the required number of samples to obtain statistical knowledge of the channel is therefore vital for \gls{urllc}. One promising option is to exploit the spatial correlation of the channel, e.g., by using \textit{radio maps}~\cite{Abou2013, Chowdappa2018, Sato2021, Wang2020, Mu2021}. These maps usually model the average \gls{snr} across space using methods such as Kriging interpolation \cite{Chowdappa2018, Sato2021} and \glspl{gp} \cite{Wang2020} and have been applied for, e.g., resource allocation \cite{Abou2013}, 
positioning \cite{Wang2020}, and route planning \cite{Mu2021}. However, because the maps considered in these works focus on the average \gls{snr}, they provide limited information about the channel distribution required for \gls{urllc} (unless a parametric model is assumed, such as Rayleigh fading). 

Modeling the \gls{snr} beyond its average is usually avoided in radio maps as the spatial correlation of fast fading decorrelates within a few wavelengths. However, the spatial independence of fast fading only applies to the instantaneous values and not its long-term statistics. In fact, it is reasonable to expect that the long-term statistics vary smoothly in space due to shared dominant paths, scatterers, etc. Thus, channel samples collected from other users in the network can, when combined with location information, be expected to be good predictors of the channel statistics beyond the average at a new location. Motivated by this, in this paper, we propose a novel approach of using \emph{statistical radio maps} to estimate the outage capacity in a wireless network. The resulting map is used to select the transmission rate for a new user joining the system, such that a specific reliability constraint can be met with high probability without the need for the device to collect any channel samples, thereby reducing the overall latency. A central feature of the proposed map is that it includes a parameter to control the trade-off between model uncertainty and transmission rate. 

The remainder of this paper is structured as follows. Sec.~\ref{sec:systemmodel_problemdef} defines the system model, and the proposed statistical map generation and rate selection scheme are presented in Sec.~\ref{sec:rateprediction}. Numerical results are presented in Sec.~\ref{sec:results}, and finally, the paper is concluded in Sec.~\ref{sec:conclusion}.

\section{System model and problem definition}\label{sec:systemmodel_problemdef}

\subsection{Signal Model}
We consider a \gls{urllc} scenario comprising a single \gls{bs} serving several \glspl{ue} within an area\footnote{The method can be applied without modifications in any finite-dimensional real space, but we focus on $\mathbb{R}^2$ for clarity and assume that the \glspl{ue} are located at the same height.} 
$\mathcal{R}\subset \mathbb{R}^2$. The \gls{bs} and the \glspl{ue} communicate both in the uplink and downlink, and the packets must be delivered with high reliability and low latency to the receiver, which precludes the possibility of estimating the instantaneous channel prior to transmission. To ease the presentation, we will assume that the transmitted signals are received without interference, but we stress that this assumption is not required as long as the distribution of the interference power is stationary. Following this, a signal $\mathbf{s}\in\mathbb{C}^n$ of length $n$ transmitted either by the \gls{bs} to a \gls{ue} or by a \gls{ue} to the \gls{bs} is received as
\begin{equation}
    \mathbf{y} = h\mathbf{s} + \mathbf{z}. \label{eq:rx_signal}
\end{equation}
Here, $h\sim p(h\cond \mbf{x}, \mbs{\phi})$ is the complex channel gain (comprising both fast and slow fading) drawn from some unknown distribution conditioned on the location of the \gls{ue} $\mbf{x}\in\mathcal{R}$ and some parameter $\mbs{\phi}$ characterizing the environment. Specifically, $\mbs{\phi}$ is common to all \glspl{ue} and captures all propagation parameters within the cell, such as the location of blockages and scatterers, providing the spatial consistency and correlation to $h\cond\mbf{x}$. Finally, $\mathbf{z}\in\mathbb{C}^n$ denotes the additive white Gaussian noise with elements drawn independently from $\mathcal{CN}(0,BN_0)$, where $N_0$ is the power spectral density and $B$ is the bandwidth. The transmitted signal is assumed to be normalized, i.e., $E[\|\mathbf{s}\|_2^2]=n$, so that the instantaneous \gls{snr} of the signal is 
\begin{equation}
    W=\frac{|h|^2}{BN_0}. \label{eq:snr}
\end{equation}
The \gls{snr} is conditionally independent given $\mbf{x}$ and $\mbs{\phi}$ across both time and users. However, the environment, characterized by $\mbs{\phi}$, will remain fixed, so we will denote the distribution of the \gls{snr} simply as $W\sim p(W\cond\mbf{x})$ for each location $\mbf{x}$. Under this model, the error probability of a packet transmitted with a given rate $R$ is dominated by the event of outage~\cite{yang14quasistatic}, which is given by the probability that the instantaneous rate supported by the channel is less than $R$, i.e.,
\begin{equation}
    p_{\text{out}}(R) = P(\log_2(1+W) < R) = F_{W}(2^{R}-1), \label{eq:Pout}
\end{equation}
where $F_{W}$ is the \gls{cdf} of $W$ given $\mbf{x}$.

\subsection{Problem Definition}
\label{sec:ProblemDef}

We consider the general problem of selecting the maximum rate $R$ for a user at any given location $\mathbf{x}$ within the cell while ensuring that the outage probability in \eqref{eq:Pout} stays below some $\epsilon \in (0,1)$. Under the assumption that the distribution of $W$ is perfectly known, this translates to selecting the rate equal to the $\epsilon$-outage capacity
\begin{align}
    R_{\epsilon} = \sup_{R}\{{R} \geq 0 \cond p_{\text{out}}(R) \leq \epsilon\}= \log_2(1+ F_{W}^{-1}(\epsilon)), \label{eq:out_capacity}
\end{align}
where $F_{W}^{-1}(\epsilon)$ is the $\epsilon$-quantile of $W$. However, in our case --- or in any realistic case --- where the distribution of the \gls{snr} is unknown and needs to be estimated, the defined $\epsilon$-outage capacity is not meaningful, and the rate selection problem becomes less trivial. 

To assist the rate selection, we assume that the \gls{bs} has collected \gls{snr} measurements over some period from previous users in the network and stores a dataset $\mathcal{D}=\{\mbf{W}_d,\mbf{x}_d\}_{d=1}^D$ comprising a set of $N$ independent \gls{snr} measurements $\mbf{W}_d=\begin{bmatrix} W_{d,1} & \dots & W_{d,N} \end{bmatrix}^\T$ from $D$ locations $\mbf{x}_d$ for $d = 1,\dots, D$. It is assumed that the user locations and propagation environment (i.e., $\mbs{\phi}$) are stationary throughout the scenario so that for each user, the \gls{snr} is only affected by multipath fading characterized by $p(W_d \cond \mbf{x}_d)$. Furthermore, the user locations are assumed to be perfectly known --- see \cite{kallehauge2022} for an analysis of how localization error can affect the reliability in scenarios like the one modeled here.

We here seek to learn a function $R \in \mathcal{F}$, where $\mathcal{F}$ is the family of rate selection functions that predict the rate for any location $\mbf{x}\in\mathcal{R}$ based on the observations $\mathcal{D}$. In stark contrast to previous uses of radio maps, we aim to account for the uncertainty of \gls{snr} estimation in the rate selection function. To that end, and inspired by the concept of \gls{pcr} introduced in~\cite{Angjelichinoski2019}, we formulate the rate selection as the following optimization problem:
\begin{subequations}\label{eq:Objective}
\begin{align} 
\maximize_{R \in \mathcal{F}} \;\;&\mathop {R(\mbf{x}\cond \mathcal{D})} \hfill \\
\text{s.t.}\;\;& P\left(p_{\text{out}}(R(\mbf{x}|\mathcal{D})) > \epsilon \right) \leq \delta.\label{eq:MetaConstraint}\hfill
\end{align}
\end{subequations}
Constraint \eqref{eq:MetaConstraint}, denoted \textit{meta-probability} $\tilde{p}_{\epsilon}$, accounts for the uncertainty in the prediction due to the data $\mathcal{D}$, and thus $\delta$ controls how much weight the system should give to model uncertainties when selecting the rate. We can observe that even though solving \eqref{eq:Objective} would render a rate meeting the reliability constraint, computing \eqref{eq:MetaConstraint} is infeasible unless the channel and data distributions --- namely $p(W\cond \mbf{x})$ and the distribution of the sampling process --- are perfectly known. Hence, in the following sections, we provide an approximate solution to the rate selection problem based on the statistical radio maps generated from $\mathcal{D}$ and the associated uncertainty.

\section{Rate Selection using Statistical Radio Map}\label{sec:rateprediction}
This section presents the framework to solve the objective defined in \eqref{eq:Objective} by modeling and predicting the spatial behavior of the channel gain $h$ using statistical radio maps. The framework is divided into three phases. First, we use the dataset $\mathcal{D}$ to obtain a non-parametric estimate of the $\epsilon$-quantile of the \gls{snr}. Then, we construct a statistical radio map that models the spatial behavior of the quantile using a Gaussian process, which provides an inherent way to capture the uncertainty in the predictions. Finally, the radio map is used to predict the rate at new locations using an estimated meta-probability that elegantly considers the model's uncertainty. Note that although we focus here on the quantile prediction, a similar approach can be used to construct a radio map that characterizes any statistic that may assist the communication. 

\subsection{Non-parametric quantile estimation} \label{subsec:modeling}
 We adopt a non-parametric estimator to estimate the quantile \gls{snr} at each location in the dataset. This contrasts with the typical approach of assuming a parametric fading distribution, e.g., Rayleigh/Rician or Nakagami fading, and then estimating the distribution parameters using e.g., maximum likelihood optimization. However, while parametric distributions typically require few samples to estimate, they are susceptible to model mismatch, which may result in prediction errors greater than what can be tolerated in \gls{urllc}~\cite{Angjelichinoski2019}.

Therefore, for each entry in the available dataset $\mathcal{D}$, we first estimate the $\epsilon$-quantile of the logarithmic-scale \gls{snr} as \cite{Ord1994} 
\begin{align}
    \widehat{q}_{\epsilon,d} = \mathcal{W}_{d,(r)}, \quad r = \lfloor N\epsilon \rfloor,\quad d = 1,\dots,D \label{eq:quantile_est}
\end{align}
where $\mathcal{W}_{d,(r)}$ is the $r$-th order statistics of $\mbs{\mathcal{W}}_d = \ln(\mbf{W}_d)$ and $\lfloor \cdot \rfloor$ is the floor function. This generates a new dataset $\mathcal{D}_\epsilon=\{ \widehat{q}_{\epsilon,d},\mbf{x}_d\}_{d=1}^D$ with the estimated $\epsilon$-quantile at observed locations on the radio map. Note that the estimate in \eqref{eq:quantile_est} is unbiased and admits an asymptotic Gaussian distribution as $N\to\infty$ \cite[p. 356]{Ord1994}. Notably, the quantile estimate does not require knowledge of the underlying distribution of the \gls{snr} and can be used as input for spatial prediction. The main disadvantage compared to parametric estimates, on the other hand, is the excessive number of samples required for estimation when $\epsilon$ is low. In fact, \eqref{eq:quantile_est} shows that $N$ scales inversely with $\epsilon$, and $\widehat{q}_{\epsilon,d}$ is only well defined when $N \geq 1/\epsilon$. However, the \gls{snr} samples used to estimate the quantiles as input to the \gls{gp} can be collected over a long period.

\subsection{Spatial interpolation with Gaussian processes} \label{subsec:GP}
We proceed to model the spatial variation of the estimated $\epsilon$-quantiles $\widehat{q}_{\epsilon,d}$ across the cell area $\mathcal{R}$, which will allow us to predict the quantile at a new location $\mbf{x}^*$ that is not contained in $\mathcal{D}$. To this end, we first normalize the quantiles as
\begin{equation}
    \widehat{q}\hspace{.5mm}'(\mbf{x}_d) = (\widehat{q}_{\epsilon,d} - \bar{q})/s, \label{eq:normalize}
\end{equation} 
where $\bar{q}=\frac{1}{D}\sum_{d=1}^D \widehat{q}_{\epsilon,d}$ and $s=\sqrt{\frac{1}{D}\sum_{d=1}^D (\widehat{q}_{\epsilon,d}-\bar{q})^2}$ are the sample mean and standard deviation of $\widehat{q}_{\epsilon,d}$, respectively. Following the result that order statistic based quantile estimates are asymptotic Gaussian \cite{Ord1994}, we assume the observation model $\widehat{q}\hspace{.5mm}'(\mbf{x}_d) = q'(\mbf{x}_d) + \xi$, where $\xi$ is an independent Gaussian random variable with zero mean and variance $\sigma_{\xi}^2$. Additionally, it is assumed that $q'$ is a Gaussian process~\cite{Rasmussen2006}, where $\mbf{q}'(\mbf{X})=\begin{bmatrix}q'({\mbf{x}_1}) & \ldots & q'({\mbf{x}_L})\end{bmatrix}^\T$ at any finite subset of locations $\mbf{X}=\begin{bmatrix}\mbf{x}_1 & \ldots & \mbf{x}_L\end{bmatrix}^\T$ is jointly Gaussian such that 
\begin{align} \label{eq:multi_gauss}
    \mbf{q}'(\mbf{X})\sim \mathcal{N}\left(
    \mbs{\mu}(\mbf{X}),
    \mbs{\Sigma}_{\mbf{X}\mbf{X}}
    \right).
\end{align}
Here, the mean vector $\mbs{\mu}(\mbf{X})\in\mathbb{R}^L$ is defined in terms of the mean function $[\mbs{\mu}(\mbf{X})]_i=m(\mbf{x}_i;\mbs{\theta}_m)$ for $i = 1,\dots,L$, parameterized by $\mbs{\theta}_m$. The elements of the covariance matrix $\mbs{\Sigma}_{\mbf{X}\mbf{X}}\in\mathbb{R}^{L\times L}$ are given by $[\mbs{\Sigma}(\mbf{X})]_{ij} = k(\mbf{x}_i,\mbf{x}_j; \mbs{\theta}_k)$, where $k(\mbf{x}_i,\mbf{x}_j; \mbs{\theta}_k)$ is a symmetric kernel function parameterized by $\mbs{\theta}_k$. In  the context of radio channels, the absolute exponential 
kernel
\begin{align}
        k(\mbf{x}_i,\mbf{x}_j;\mbs{\theta}_k) = \sigma_k^2\exp\left(-\frac{\norm{\mbf{x}_i - \mbf{x}_j}_2}{d_c}\right) \label{eq:expabs}
\end{align}
with parameters $\mbs{\theta}_k=(\sigma_k^2,d_c)$, is referred to as the Gudmundson correlation model~\cite{Gudmundson1991}, and has been widely applied along with a log-distance mean function to model how the average \gls{snr} varies across space~\cite{Chowdappa2018,Sato2021}. Note that $\sigma_k^2$ is the variance of the Gaussian process and $d_c$ is the \textit{correlation distance}, which tends to be in the order of the size of blocking objects when modeling shadow fading \cite{Goldsmith2005}. Through numerical experimentation, we found that the Gudmundson correlation model is also well suited to model the $\epsilon$-quantile of the \gls{snr} which is therefore adopted  to characterize the spatial correlation of $\mbf{q}'$. For the mean, we use $m(\mbf{x}_i) = 0$, as commonly done in the literature~\cite{Rasmussen2006}.

The radio map is constructed for a regular grid of $L$ locations simultaneously, denoted $\mbf{X}^* = \begin{bmatrix}
\mbf{x}_1^* & \dots & \mbf{x}_L^*
\end{bmatrix}^\T$. In order to predict the Gaussian process at locations $\mbf{X}^*$, we express the joint distribution of the noisy observations $\widehat{\mbf{q}}'(\mbf{X})\in\mathbb{R}^D$ and quantiles in the grid $\mbf{q}'(\mbf{X}^*) \in \mathbb{R}^{L}$ as
\begin{align} \label{eq:joint_gauss}
    \begin{bmatrix*}[l]
    \mbf{q}'(\mbf{X}^*) \\
    \widehat{\mbf{q}}'(\mbf{X})
    \end{bmatrix*} \sim \mathcal{N}\left(
    \mbf{0},
    \begin{bmatrix*}[c]
    \mbs{\Sigma}_{\mbf{X}^*\mbf{X}^*} & \mbs{\Sigma}_{\mbf{X}^* \mbf{X}} \\
    \mbs{\Sigma}_{\mbf{X}\mbf{X}^*} & \mbf{\Sigma}_{\mbf{X}\mbf{X}}+\sigma_{\xi}^2\mbf{I}_{D}
    \end{bmatrix*}
    \right),
\end{align}
where $\mbf{I}_D\in\mathbb{R}^{D\times D}$ is the identity matrix. Following \eqref{eq:joint_gauss}, the predictive distribution for $\mbf{q}'(\mbf{X}^*) \cond \mbs{\vartheta}$ where $\mbs{\vartheta} = \left(\widehat{\mbf{q}}'(\mbf{X}), \mbf{X}, \mbf{X}^*,  \mbs{\theta}_k\right)$ is also a multivariate Gaussian distribution with \cite{Rasmussen2006}
\begin{align}
    E[\mbf{q}'(\mbf{X}^*) \cond \mbs{\vartheta}] &= \mbs{\Sigma}_{\mbf{X}^* \mbf{X}}(\mbs{\Sigma}_{\mbf{X}\mbf{X}}+\sigma_{\xi}^2\mbf{I}_{D})^{-1}\widehat{\mbf{q}}'(\mbf{X}), \label{eq:pred_mean} \\
    \Cov[\mbf{q}'(\mbf{X}^*) \cond \mbs{\vartheta}] &= \mbs{\Sigma}_{\mbf{X}^*\mbf{X}^*}- \mbs{\Sigma}_{\mbf{X}^*\mbf{X}}(\mbs{\Sigma}_{\mbf{X}\mbf{X}}+\sigma_{\xi}^2\mbf{I}_{D})^{-1}\mbs{\Sigma}_{\mbf{X}\mbf{X}^*}. \label{eq:pred_cov}
\end{align}
Equations \eqref{eq:pred_mean} and \eqref{eq:pred_cov} constitute the predictive distribution for $\mbf{q}'(\mbf{X}^*)$ and are referred to as the \textit{predictive mean} and \textit{covariance}, respectively. Because the model produces a full distribution of the predicted quantiles, it allows us to capture the uncertainty of the estimates, which we will make use of in the rate selection in the next subsection.

Finally, we note that so far we have assumed that the hyperparameter $\mbs{\theta}_k$ is fixed. However, in practice it needs to be estimated as well. We do this using maximum likelihood estimation by numerically maximizing the likelihood of the observed data $\widehat{\mbf{q}}'(\mbf{X})$ with respect to $\mbs{\theta} = (\sigma_k^2, d_c, \sigma_{\xi}^2)$. We gently refer the reader to \cite[Ch. 5]{Rasmussen2006} for further details on the procedure.

\subsection{Rate selection function}

Once the statistical radio map is generated at the \gls{bs} and broadcasted to new \glspl{ue} joining the system at location $\mbf{x}^*_l$, the predictive distribution of the quantile can be used to select the rate. 
Specifically, we aim to use the statistical information about $q'(\mbf{x}^*_l)$ to solve \eqref{eq:Objective}. Note that we assume $\mbf{x}^*_l$ to be known.

As discussed in Sec. \ref{sec:ProblemDef}, the difficulty of computing the meta-probability $\tilde{p}_\epsilon$ in \eqref{eq:MetaConstraint} is the requirement of perfect knowledge of both channel and data distributions, which is unfeasible in any realistic setup. To circumvent this, we replace this knowledge by the predictive distribution of the quantile $q'(\mbf{x}^*_l)$ from the statistical map, rendering an approximated solution to \eqref{eq:Objective}. From \eqref{eq:Pout} and \eqref{eq:MetaConstraint}, we can write
\begin{align}
    \tilde{p}_{\epsilon} &= P\left(P(R(\mbf{x}^*_l \cond \mathcal{D}) > \log_2(1+W_l)) > \epsilon \right) \notag \\
    &= P(P(\ln(2^{R(\mbf{x}^*_l\cond\mathcal{D})}-1) > \ln(W_l) ) > \epsilon), \label{eq:Meta2}
\end{align}
\newgeometry{top=.73in,bottom=0.95in, inner = 0.7in, outer = 0.7in}
where $W_l$ is the \gls{snr} at location $\mbf{x}^*_l$. Denoting $F_{\mathcal{W}_l} (\mathcal{W}_l \cond \mbf{x}^*_l)$ as the \gls{cdf} of $\mathcal{W}_l = \ln(W_l)$, we have
\begin{equation}
     \tilde{p}_{\epsilon}=P(\ln(2^{R(\mbf{x}^*_l \cond \mathcal{D})}-1) > F_{\mathcal{W}_l}^{-1}(\epsilon\cond\mbf{x}^*_l)),
\end{equation}
which is directly given by the \gls{cdf} $F_{q_\epsilon}(q\cond\mbf{x}^*_l)$ of the true \gls{snr} quantile (in logarithmic scale), i.e.,
\begin{equation}
     \tilde{p}_{\epsilon}=F_{q_{\epsilon}}(\ln(2^{R(\mbf{x}^*_l\cond\mathcal{D})}-1)\cond\mbf{x}^*_l).
\end{equation}

Since $F_{q_\epsilon}(\cdot)$ is unknown, we replace it by the predictive distribution obtained from the statistical radio map, from which we have the normalized quantile $q'(\mbf{x}^*_l)\sim\mathcal{N}(\mu_{q,l}, \sigma_{q',l}^2)$, with $\mu_{q',l}$ and $\sigma_{q',l}^2$ as in \eqref{eq:pred_mean} and \eqref{eq:pred_cov}. Then, we denormalize the quantile by isolating from \eqref{eq:normalize}, leading to $q_{\epsilon,l}^*\sim\mathcal{N}(\mu_l, \sigma_l^2)$ with $\mu_l = s\mu_{q',l} + \overline{q}$ and $\sigma_l^2 = \sigma_{q',l}^2 s^2$, where $\bar{q}$ and $s$ are the global mean and standard deviation obtained from the dataset as defined in Sec.~\ref{subsec:GP}. Hence, given the meta-probability constraint $\delta$, an approximate solution to \eqref{eq:Objective} fulfills
\begin{align}
F_{q_{\epsilon,l}^*}(\ln(2^{R(\mbf{x}^*_l\cond\mathcal{D})}-1)\cond\mbf{x}^*_l) = \delta,
\end{align}
whose solution follows as 
\begin{align}
    R(\mbf{x}^*_l&\cond\mathcal{D}) = \log_2\left(1 + \exp\left(F_{q^*_{\epsilon},l}^{-1}(\delta)\right)\right) \nonumber \\
    &= \log_2\left(1 + \exp\left(\mu_l + \sqrt{2}\sigma_l\mathrm{erf}^{-1}(2\delta - 1)\right)\right), \label{eq:rate_select}
\end{align}
where $\mathrm{erf}^{-1}$ is the inverse error function. Note that the meta-probability only matches the target confidence when the modeling assumptions for quantile estimation and spatial correlation are correct. Nevertheless, $\delta$ provides a way to control how conservative the selected rate should be. The entire procedure (map generation and rate selection) is summarized in Algorithm~\ref{alg:statmap}.

\begin{algorithm}
\small
\caption{Rate selection via statistical radio map }\label{alg:statmap}
\begin{algorithmic}[1]
\REQUIRE \gls{snr} and locations $\mathcal{D}=\{\mbf{W}_d,\mbf{x}_d\}_{d=1}^D$, new locations for prediction $\mbf{X}^*$.
\STATE \textbf{Estimate} $\widehat{q}\hspace{.5mm}'(\mbf{x}_d)$ for $d = 1,\dots,D$ using \eqref{eq:quantile_est} and \eqref{eq:normalize}. 
\STATE \textbf{Estimate} $\widehat{\mbs{\theta}}_k = \argmax_{\mbs{\theta}_k} f(\widehat{\mbf{q}}'(\mbf{X}); \mbs{\theta}_k)$, $f$ is the joint likelihood function of $\widehat{\mbf{q}}'(\mbf{X})$. 
\STATE \textbf{Compute} the covariances from \eqref{eq:joint_gauss} based on $\widehat{\mbs{\theta}}_k$ and kernel $k$.
\STATE \textbf{Compute} parameters for the predictive distribution of $\mbf{q}(\mbf{X}^*) \cond \mbs{\vartheta}$ according to \eqref{eq:pred_mean} and \eqref{eq:pred_cov}
\FORALL{$\mbf{x}^*_l \in \mbf{X}^*$}
\STATE \textbf{Compute} $\mu_l = s\mu_{q',l} + \overline{q}$ and $\sigma_l^2 = \sigma_{q',l}^2 s^2$
\STATE \textbf{Select} $R(\mbf{x}^*_l\cond\mathcal{D})$ using \eqref{eq:rate_select}
\ENDFOR
\end{algorithmic}
\end{algorithm}

\section{Numerical evaluation} \label{sec:results}

For numerical evaluation, we consider \glspl{ue} in the area $\mathcal{R} = [-50,50]\times [-50,50]$ m with the \gls{bs} in the middle at $\mbf{x}_{\text{bs}} = \mbf{0}$ m. The dataset $\mathcal{D}$ is composed of $D=500$ locations (unless otherwise stated), each containing $N=10^5$ independent \gls{snr} observations. To imitate a realistic clustered setup, the locations are sampled from a modified Thomas process \cite{moller2003statistical}. We first simulate a continuous Thomas process in the area $[-71,71]\times [-71,71]$ (to remove edge effects) with a parent process rate of $0.005$, a daughter process rate of $100$, and a standard deviation of $3.5$. Points outside $\mathcal{R}$ are discarded, and the remaining points are rounded to the nearest two meters and then randomly and independently thinned until $D$ unique points remain.

The channel coefficients $h$ used to compute the \glspl{snr} in \eqref{eq:snr} across the region are simulated with the simulation tool QuaDRiGa using the 3GPP NR Urban Micro-Cell scenario with line of sight (see \cite[p. 81]{quadriga}) with \gls{bs} height $10$ m, \gls{ue} height $1.5$ m, transmit power of $0$ dBm, and $2.6$ GHz as central frequency. Since QuaDRiGa is a geometric channel simulator, we emulate independent fast fading by adding uniformly distributed random phase shifts 
on top of each path arriving to the receiver, and then adding up the contributions of all the paths to obtain a narrowband channel coefficient. With the resulting channel gains, the \gls{snr} values are calculated as in \eqref{eq:snr} with $B=200$ KHz and noise power $BN_0 = -115$ dBm. To evaluate the predicted outage probabilities, we also simulate test data $\mathcal{D}_\text{test}$ comprising $N = 10^5$ \gls{snr} values at $D_\text{test}= 2601$ locations forming a uniform grid with $2$ m spacing.

\begin{figure}
    \centering
    \vspace{.14in}
    \includegraphics[width = \linewidth]{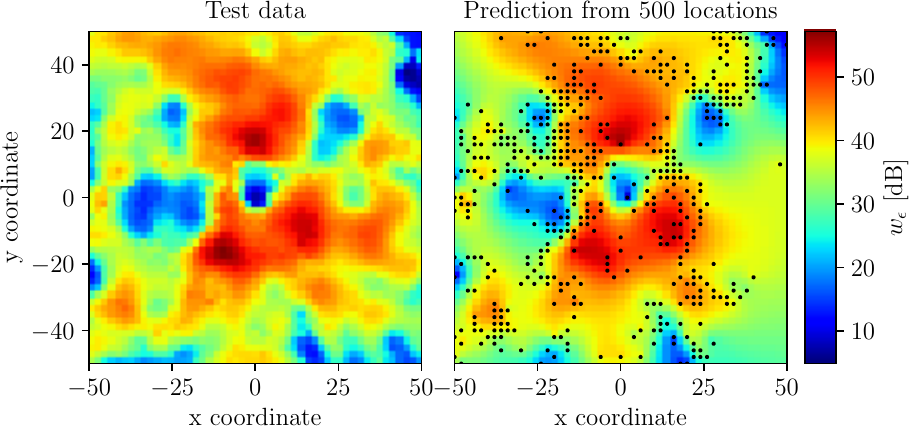}
    \vspace{-.2in}
      \caption{Predictions of the $\epsilon$-quantile $w_{\epsilon} = F_W^{-1}(\epsilon)$ for the \gls{snr} with $\epsilon = 10^{-3}$. The left plot shows the test data; the right plot shows the mean values predicted by the Gaussian process based on $\mathcal{D}$. The black dots show the locations in $\mathcal{D}$.}
    \label{fig:interpolate}
\end{figure}

We start by comparing the predictive means of the $\epsilon$-quantile of $W$ obtained for a sample dataset $\mathcal{D}$ using Algorithm \ref{alg:statmap} to the quantiles estimated from $\mathcal{D}_\text{test}$ in Fig.~\ref{fig:interpolate}. The left map shows $w_{\epsilon} = F_{W}^{-1}(\epsilon)$ obtained from $\mathcal{D}_\text{test}$, while the right map shows the quantile predicted from the radio map, i.e., $e^{\mu_i}$ for each $\mbf{x}_i\in\mathcal{D}_\text{test}$ calculated using the procedure in Sec.~\ref{sec:rateprediction}. It can be seen that the predictive mean is generally close to the test values except for areas far away from any observed location, e.g., in the right part of the map. However, the predictive variance (which is not depicted) is also significantly higher in these areas as well.

\begin{figure}
    \centering
    \includegraphics[width = \linewidth]{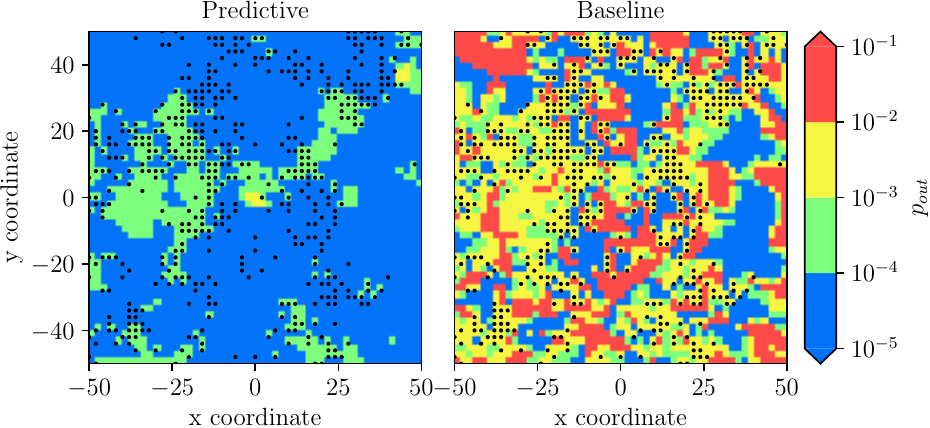}
      \caption{Outage probabilities resulting from the predictive and the baseline rate selection methods. The black dots mark the observed locations. Outage probabilities in the blue/green areas are below the target of $\epsilon = 10^{-3}$ and yellow/red areas are above.}
    \label{fig:onemap}
\end{figure}

We now turn our attention to the rate selection problem, where we aim to predict the maximum rate for the unobserved locations on the map satisfying an outage probability of at most $\epsilon = 10^{-3}$ (we refer to this as \textit{predictive} rate selection). Unless otherwise stated, we set $\delta = 10^{-3}$. For comparison, we use a baseline scheme, which simply selects the rate based on measurements from the closest observed location. Specifically, if 
$\mbf{x}_d$ is the closest observed location to the target location $\mbf{x}^*$, then the baseline scheme chooses the rate as
\newpage
\restoregeometry
\noindent $R(\mbf{x}^* \cond \mathcal{D}) = \log_2(1 + W_{d,(\lfloor n\epsilon\rfloor)})$. The outage probabilities resulting from the rate selection schemes are empirically computed using the test data --- see Fig. \ref{fig:onemap} for an example based on the same scenario as in Fig. \ref{fig:interpolate}. The figure shows that the outage probabilities obtained using the predictive rate selection rarely exceed the target $\epsilon$, whereas the baseline scheme often exceeds the target outage probability.

\begin{figure}
    \centering
    \includegraphics[width = \linewidth]{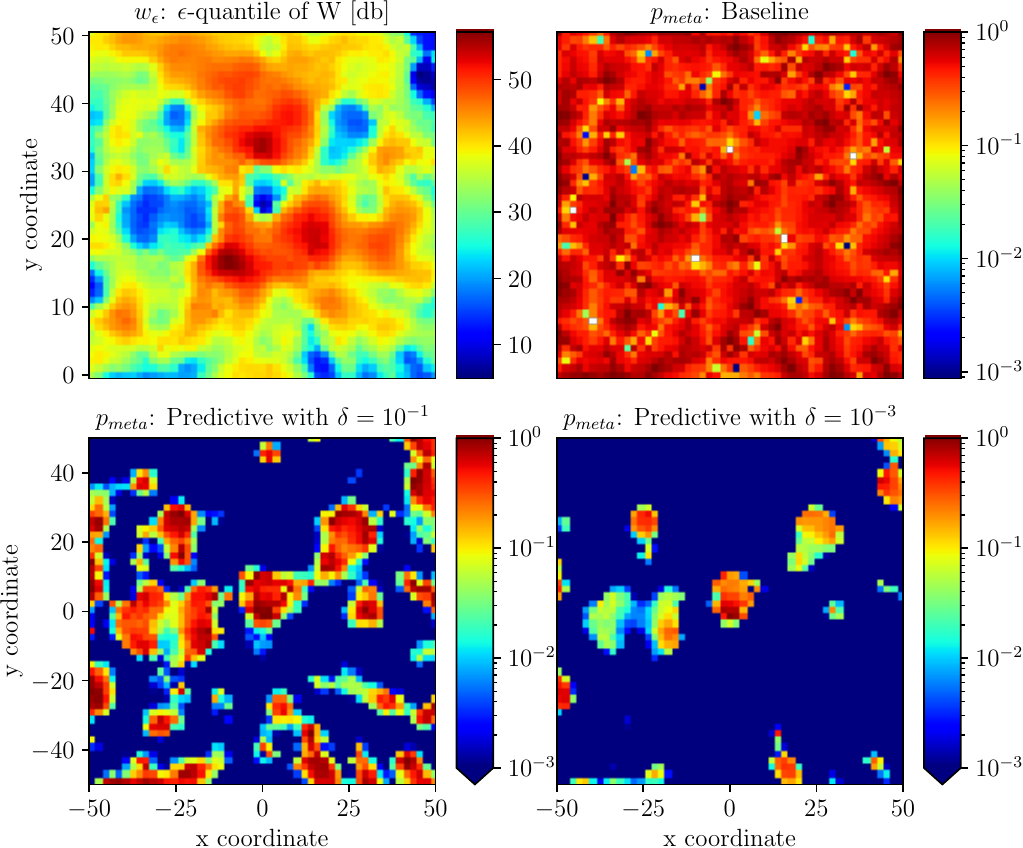}
      \caption{Top left: $\epsilon$-quantile for the \gls{snr} with $\epsilon = 10^{-3}$. Top right and lower: Estimated meta-probability from $10^4$ realizations of the \gls{ue} sampling process. The predictive rate selection is shown with two different choices of delta, i.e., $\delta = 10^{-1}$ and $\delta = 10^{-3}$.}
    \label{fig:mapaverage}
\end{figure}

The results in Figs. \ref{fig:interpolate} and \ref{fig:onemap} clearly depend on the spatial distribution of the user locations. For example, the predictive rate selection scheme is generally very conservative in areas with no observations (due to higher predictive variance), leading to an outage probability several orders of magnitude below the target. On the other hand, the baseline scheme is generally close to the target outage probability only for points close to an observed location. To analyze the performance with less dependency on the particular realization of the \gls{ue} locations in $\mathcal{D}$, we repeat the rate selection process in Algorithm \ref{alg:statmap} with the same \gls{snr} quantiles as shown top left in Fig. \ref{fig:mapaverage}, but for $10^4$ different realizations of the \gls{ue} location sampling process. We then estimate the meta-probabilities defined in \eqref{eq:Meta2} by counting the number of times the resulting outage probability exceeds $\epsilon$ at each location, as seen in the top right and lower plots in Fig. \ref{fig:mapaverage}. We see that the meta-probabilities of the baseline scheme are often above $50\%$, which means that the outage probability exceeds $10^{-3}$ for more than half of the simulations at those locations. The predictive scheme, on the other hand, is typically below the target outage probability. Interestingly, we see a strong negative correlation between areas with high meta-probability and areas where the $\epsilon$-quantile of the \gls{snr} is lower than its surroundings. For example, the predictive rate selection has a high meta-probability close to the \gls{bs} in the center, where the quantile is about $40$ dB lower than its immediate surroundings. Intuitively, this makes sense since predictions made from around the \gls{bs}, where the quantile is high, will predict a similarly high quantile, which will cause a high outage probability. A reverse effect is seen for areas with a higher quantile than its surroundings, where the outage is below the target with high probability. 

\begin{figure}
    \centering
    \includegraphics[width = \linewidth]{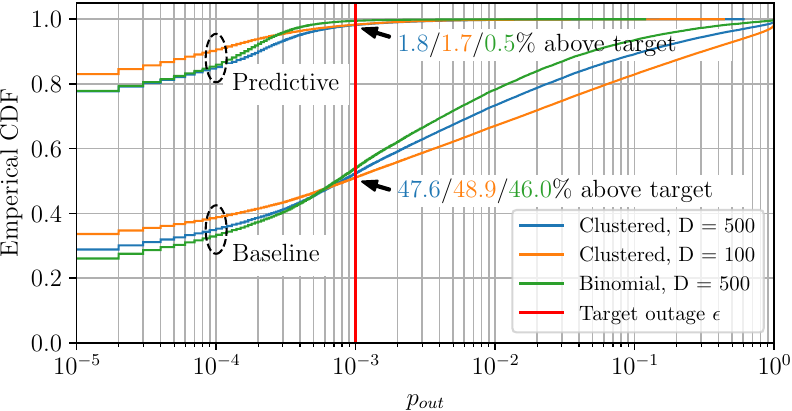}
    \caption{Emperical \gls{cdf} of outage probabilities for different scenarios.}
    \label{fig:CDF}
\end{figure}

Fig.~\ref{fig:CDF} shows the empirical \gls{cdf} of the outage probabilities $p_{\text{out}}$ obtained across all locations on the map and $10^4$ realizations of the \gls{ue} locations in $\mathcal{D}$. 
The figure also shows the distribution of outage probabilities for two new scenarios, namely when rate selection is based only on observations from $D = 100$ locations, and when the \gls{ue} locations in $\mathcal{D}$ are drawn according to a binomial point process, i.e., uniformly from $\mathcal{R}$.

Similarly to the previous results, we observe that the outage probabilities are generally lower for the predictive rate selection than the baseline. The probabilities at $p_{\text{out}} = 10^{-3}$ correspond to $1 - \tilde{p}_{\epsilon}$, and we see that the predictive rate selection slightly exceeds the target meta-probability of $\delta = 0.1\%$ in the three scenarios, which is due to modeling mismatches. However, it is significantly closer to the target than the baseline scheme. The meta-probability is closer to the target reliability in the scenario with the binomial point process than the clustered process, which is expected since the test locations are also located uniformly across the map. Furthermore, the outage probability for predictive rate selection has a high probability not only of being below the target of $10^{-3}$ but also of being below $10^{-5}$ or even lower, which reflects the high cost of being uncertain about the channel statistics.

\begin{figure}
    \centering
    \includegraphics[width = \linewidth]{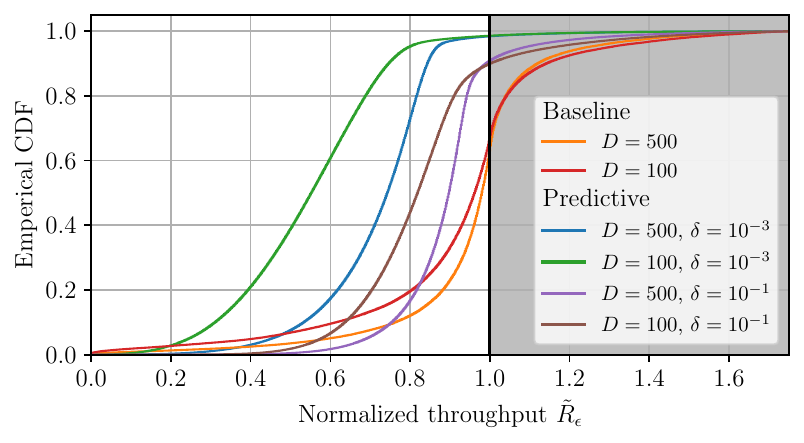}
    \caption{Empirical \gls{cdf} of the normalized throughput for $D = 100$ and $500$ observations and with $\delta = 10^{-3}$ and $10^{-1}$. A value greater than 1 violates the reliability constraint.}
    \label{fig:throughput}
\end{figure}

To evaluate this aspect, we define the throughput resulting from rate $R$ as the average rate when the system is not in outage \cite{Angjelichinoski2019}, i.e.,
\begin{align}
    E_W[R \, \mathbbm{1}\{R \leq \log_2(1+W)\}] = R(1-p_{\text{out}}(R))
\end{align}
where $\mathbbm{1}$ is the indicator function. To account for spatial variation of the \gls{snr}, we compute the normalized throughput, denoted $\tilde{R}_{\epsilon}$, by dividing with the throughput of the $\epsilon$-outage capacity $R_{\epsilon}$ from \eqref{eq:out_capacity} such that
\begin{align}
    \tilde{R}_{\epsilon} =  \frac{R(1-p_{\text{out}}(R))}{R_{\epsilon}(1-\epsilon)}. 
\end{align}
Note that from this definition, a value $\tilde{R}_{\epsilon}>1$ means that the rate is greater than the $\epsilon$-outage capacity, and thus the target reliability constraint is not satisfied. Fig.~\ref{fig:throughput} shows the distribution of the normalized throughput resulting from the considered rate selection methods. It can be seen that the throughput for the predictive rate selection in far most cases is smaller than one, meaning that the reliability constraint is satisfied. This is in contrast to the baseline scheme, which in many cases does not meet the target reliability. Furthermore, although the throughput of the predictive rate selection is conservative, it is generally far from zero, suggesting that the proposed method is suitable for rate selection in the \gls{urllc} regime.

\section{Conclusion}\label{sec:conclusion}
In this paper, we have considered the problem of rate selection for \gls{urllc} when the channel statistics are unknown. To circumvent the problem of having to collect a very large number of samples for non-parametric channel estimation, we have proposed a framework that unites the ideas of radio maps and statistical learning. In particular, our framework uses a \gls{gp} radio map to model the distribution of the $\epsilon$-outage capacity across space, allowing us to predict the outage capacity at unseen locations using non-parametric quantile estimates obtained from only a few locations. Using the predicted distribution, we introduced a rate selection rule inspired by the concept of meta-probabilities from statistical learning, which controls how prediction uncertainty penalizes the rate. Through numerical results, we demonstrated that the proposed framework can accurately model the outage capacity and that the rate selection rule satisfies the reliability target with significantly higher probability than a baseline scheme, confirming that the framework is suitable for \gls{urllc}. Future work will include extensions such as the ability to handle non-stationary propagation environments, modeling of localization uncertainty, and reducing the number of measurements required at each location.

\bibliographystyle{IEEEtran}
\bibliography{references}   

\end{document}